\documentclass[conference]{IEEEtran}
\IEEEoverridecommandlockouts
\usepackage{cite}
\usepackage{amsmath,amssymb,amsfonts}
\usepackage{algorithmic}
\usepackage{graphicx}
\usepackage{textcomp}
\usepackage{xcolor}
\usepackage[mode=match]{siunitx}
\usepackage[normalem]{ulem}

\def\BibTeX{{\rm B\kern-.05em{\sc i\kern-.025em b}\kern-.08em
    T\kern-.1667em\lower.7ex\hbox{E}\kern-.125emX}}

\begin{document}

\title{OH$^-$ ions can reduce the iodide migration in MAPI} 

\author{
\IEEEauthorblockN{
\textbf{R. E. Brophy}\IEEEauthorrefmark{1},
\textbf{M. Kateb}\IEEEauthorrefmark{2},
\textbf{I. Ghitiu}\IEEEauthorrefmark{3}\IEEEauthorrefmark{4},
\textbf{N. Filipoiu}\IEEEauthorrefmark{3}\IEEEauthorrefmark{6},
\textbf{K. Torfason}\IEEEauthorrefmark{1},
\textbf{H. G. Svavarsson}\IEEEauthorrefmark{1},\\
\textbf{G. A. Nemnes}\IEEEauthorrefmark{3}\IEEEauthorrefmark{5}\IEEEauthorrefmark{6},
\textbf{I. Pintilie}\IEEEauthorrefmark{7},
\textbf{A. Manolescu}\IEEEauthorrefmark{1}
}\\

\IEEEauthorblockA{Email: Rachel18@ru.is\\
{\IEEEauthorrefmark{1}Department of Engineering, Reykjavik University, Menntavegur 1, IS-102 Reykjavik, Iceland}\\
{\IEEEauthorrefmark{2}%Condensed Matter \& Materials Theory Division, 
Department of Physics, Chalmers University of Technology, SE-412 96 Gothenburg, Sweden}\\
{\IEEEauthorrefmark{3}University of Bucharest, Faculty of Physics, 077125 Magurele-Ilfov, Romania}\\
{\IEEEauthorrefmark{4}National Institute for Laser, Plasma and Radiation Physics, 077125 Magurele-Ilfov, Romania}\\
{\IEEEauthorrefmark{5}Research Institute of the University of Bucharest (ICUB), Mihail Kogalniceanu Blvd 36-46, 050107 Bucharest, Romania}\\
{\IEEEauthorrefmark{6}Horia Hulubei National Institute for Physics \& Engineering, 077126 Magurele-Ilfov, Romania}\\
{\IEEEauthorrefmark{7}National Institute of Materials Physics, 077125 Magurele-Ilfov, Romania}\\
}
}

\maketitle

\begin{abstract}
One of the main degradation mechanisms of methylammonium lead iodine (MAPI), which is an important material for perovskite based solar cells, is the migration of iodide ions. It is believed that this phenomenon is in fact dominated by the diffusion of iodide vacancies. In this paper, we suggest that the addition of a small amount of OH$^-$ ions can help suppress the migration of iodide and increase the overall stability of the material.  Through the use of molecular dynamics simulations, we show that the OH$^-$ ions can bind to the positively charged iodide vacancies and can block the access of the negative iodide ions into those vacancies.
\end{abstract}

\vspace{3mm}


\section{Introduction}
The hybrid organic-inorganic perovskite materials, and in particular the methylammonium lead iodine (CH$_3$NH$_3$PbI$_3$, or MAPI), have become a significant topic of interest in the last decade due to their photovoltaic properties. Since 2013, perovskite based solar cells have increased in efficiency from 13\% to roughly 25\%, making this category one of the fastest-growing photovoltaic devices. The main interest in the perovskite material is that it is able to convert light into electricity within its natural structure, significantly lowering the manufacturing costs \cite{OIHP_book,Elsa_book}. 
Also adding to the lower manufacturing cost is that it only requires modest fabrication equipment, much simpler than needed for conventional silicon-based solar cells. However, the perovskite based solar cells have the disadvantage of a limited stability. One main reason for that is the degradation caused by ionic migration within the perovskite material, especially the diffusion of iodide \cite{Eames2015, Besleaga2016}. Another consequence is a long relaxation time of the photo-induced current leading to a hysteresis of the current-voltage characteristic when the voltage is varied in time \cite{Sanchez14,Tress2015,Nemnes17a,Filipoiu2022}.  

Quantitative data about the dynamics of the iodide diffusion is difficult to extract from experiments since it depends on several factors, such as temperature, boundaries, various types of defects, intrinsic electric field, etc. Nevertheless, molecular dynamics (MD) simulations have shown that, at least in the homogeneous MAPI material, the dominant diffusion mechanism is the migration of the iodide vacancies, with an estimated diffusion coefficient between $2.5 - 4.3 \times 10^{-6} \ {\rm cm}^2/{\rm s}$, whereas the diffusion coefficient of iodide interstitial ions is smaller by one order of magnitude or more \cite{Delugas2016,Brophy2023}.     Hence, since the iodide ion is negative, once an iodide vacancy is formed, it will be positively charged, and a nearby iodide will migrate toward it and attempt to fill it up, and simultaneously creating a new vacancy.  Therefore, since the vacancy appears to move faster than the iodide ion, it is tempting to think that blocking the vacancy would reduce the overall diffusion of the iodide.

Another known cause of degradation of the perovskite-based solar cells is the introduction of humidity (water) into the device. Water molecules can have multiple effects.  Depending on the amount of H$_2$O molecules, the absorption factor may significantly decrease \cite{Wang2016}, iodide ions may be dislocated \cite{Mosconi2015}, and other detrimental phenomena may occur. Still, a small amount of water can be beneficial for the photovoltaic properties, for example by delaying the electron-hole recombination events \cite{Long2016}, with negligible changes of the band gap \cite{Mosconi2015}. 

%However, to our knowledge, there has yet to be either theoretical or experimental data for the introduction of just OH- into the perovskite material. 

Atomistic calculations based on the density functional theory (DFT) have shown that the dissociation of a water molecule into H$^+$ and OH$^-$ ions in MAPI requires an energy of 0.75~eV in dark, and 0.57~eV in the presence of photo generated electrons \cite{Peng2018}.  Therefore, if a small amount of water is present in the light-harvesting perovskite  material, we can assume the existence of some OH$^-$ (hydroxyl) ions therein.  One possibility can be that the hydroxyl is generated at the surface and enters in MAPI along grain boundaries. Another possibility is a controlled adsorption during the preparation of MAPI.  Other ways to obtain OH$^-$ in MAPI may also exist. 

In this paper we use MD simulations to show that an OH$^-$ interstitial can reduce the migration of the iodide ions inside MAPI, by binding to an iodide vacancy, and blocking the path of the iodide ions through that site.  We have also tested, using density functional theory (DFT) based calculations, that the presence of hydroxyl ions at the interfaces of MAPI with the electron and hole transporter materials (TiO$_2$ and Cu$_2$O, respectively) does not have a significant effect on the energy gaps and band alignment. %\\

\section{Methods}

We utilize MD simulations to describe the movement of all atoms in the MAPI system.  The simulations were performed using the LAMMPS software package \cite{Plimpton1995} which solves Newton's equations of motion for each atom or molecule, based on interatomic effective forces. We used the perovskite structure with CH$_3$NH$_3$PbI$_3$ stoichiometry in which the methylammonium (MA) molecule has a 1:1 ratio with Pb, and is located in the middle of a cubic cell (or cage) made of PbI$_3$. 

To describe the interatomic forces we used the MYP potential derived by Mattoni et al. \cite{Mattoni2015}. This potential considers bonded interactions in MA and non-bonded for the PbI cage and PbI--MA link. 
%The MYP potential considers both bonded interactions with MA, as well as any non-bonded interactions. This potential allows for the computation of movement for all iodide, lead, and MA constituents.  %Briefly, the MYP considers bonded interactions in MA and non-bonded PbI cage and PbI--MA. 
This combination allows the computation of the local movement or diffusion of all iodide, lead, and MA constituents.  The random rotation of the MA molecule at 300~K gives a cubic phase. The intra-MA interactions were modeled using a bounded generalized AMBER force field (GAFF), and the Buckingham force field was used for Pb-Pb, I-I, Pb-I, and Pb/I-C/N interactions Pb/I-H  interactions were modeled with Lennard-Jones potential. For the hydroxyl we used TIP3P parameters with charges -0.8476 and 0.4238~$e$  on O and H, respectively. This gives the hydroxyl group a total -0.4238~$e$ as compared to +1.13~$e$ for iodide vacancy. 

The simulated system was made of 256 unit cells (MAPbI$_3$), with a total of 3072 atoms, which were relaxed by energy minimization followed by ramping up and maintaining its temperature at 300~K and the constant pressure of 1~bar for 1 ns. This produces samples from an isothermal-isobaric (NPT) ensemble controlled by the Nose-Hoover thermostat and barostat.  Once the system was relaxed, a single iodide was removed, theoretically making the system positively charged. However, due to the large size of the system, there are no artificial implications on the simulations. To see if there was a way to mediate the migration, the hydroxyl group was placed within the system at varying positions and distances from the initial vacancy, and each case was run as a separate simulation. The positions of all iodine atoms in the NPT ensemble as function of time were recorded during each simulation. Further justification on the choice of the ensemble and other calculation details can be found elsewhere \cite{Brophy2023}.
The average computation time was about 22 CPU hours for each nanosecond on a high-performance computer cluster of AMD EPYC processors 2300 MHz.

DFT based simulations were also performed to test the impact of OH substitutions on the band gap and band alignment at the interfaces between MAPI and hole and electron transporter materials, respectively. They were carried out using the SIESTA software~\cite{Soler_2002}, a code that employs the local density approximation with the inclusion of Hubbard correction (LDA+U) to treat the electron-electron interactions, and numerical atomic orbitals as a basis set to solve the Kohn-Sham equations. Since this ab-initio method operates with atoms, and not with ions like the force based method of LAMMPS, neutral OH groups were used to substitute random iodine atoms in the MAPI layer. 

In the case of the hole transporter, the simulation cell consisted of 218 atoms, grouped in 2x2x6 unit cells of Cu$_2$O, and 1x1x3 tetragonal MAPI. The Brillouin zone was sampled using a Monkhorst-Pack k-point mesh of $7\times 7\times1$.
%and the forces on each atom were relaxed until they were less than 0.04 eV/\AA. 
The system was then allowed to relax until the forces on each atom were less than 0.04 eV/\AA. The resulting optimized structures were further
%from MD simulations 
utilized to determine the electronic states.  
%The band structures were computed using the DFT-LDA+U method, and 
The partial densities of states (PDOS) were analyzed to determine the band gap and band alignment, 
%The PDOS were calculated 
using the same Monkhorst-Pack k-point mesh. %used during the relaxation,
%MD simulations, 
The energy cutoff for the basis set was 100 Ry. 

A similar computational setup was used to check the influence of OH substitutions of iodine atoms at the interface between the MAPI and a TiO$_2$ layer which has the role of electron transporter in the solar cell device.   This time the interface contained 198 atoms, and it was composed from a rutile TiO$_2$ layer, obtained by multiplying the unit cell with 2x2x5, and a MAPI 1x1x3 tetragonal layer. \\
% We have the same MAPI in Cu2O and TiO2 (I had the impression that the phase is orthorhombic) NF

\section{Results}

We computed with LAMMPS the mean square displacement (MSD) of all iodide ions included in the simulation cell, as a function of the time $t$, 
\begin{equation}
{\rm MSD}(t)=\frac{1}{N}\sum_{i=1}^{N} |{\bf r}_i(t) - {\bf r}_i(0)|^2 \ ,
\label{msddef}
\end{equation}
with $N$ being the total number of iodides and ${\bf r}_i$ the position of the particular ion $i$. The diffusion coefficient associated with the mobile ions can be obtained as 
\begin{equation}
D = \frac{1}{6}\lim_{t\to \infty} \frac{ {\rm MSD}(t) } {t} \, ,
\label{diffcoeff}
\end{equation}
which in practice is calculated using the slope of the MSD vs. time, divided by the double of the spatial dimension \cite{Allen1989}.

\begin{figure}[hbt!]
    \centering
    \includegraphics[width=1.0\linewidth]{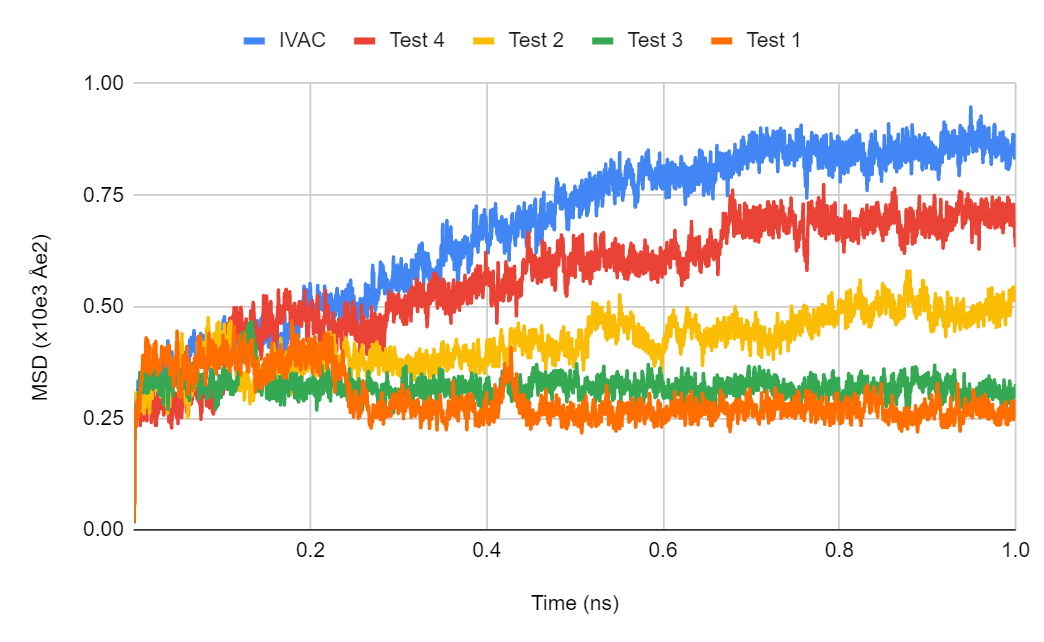}
    \vspace*{-8mm}
    \caption{MSD of all atoms at 300K for \qty{1}{\ns}. From ascending order, each test represents the vacancy and OH$^-$ group are initially placed further apart. In Test 1 they are placed within the same unit cell, in Test 2 they placed one unit cell apart, and so on.  The IVAC case corresponds to the MSD of all atoms without the addition of the OH$^-$ group.}
    \label{fig:msd_Ivac}
\end{figure}

The numerical results of the MSD are shown in Figure~\ref{fig:msd_Ivac}. We first consider a simulation cell that in the initial state includes the reference case of an iodide vacancy (IVAC) and no hydroxyl, at room temperature and normal pressure. The corresponding MSD is shown in blue, and the slope is as expected from previous studies \cite{Delugas2016,Brophy2023}. Then, in Test 1 we repeat the simulation by adding an OH$^-$ group in the same unit cell as the IVAC. Further, in Tests 2, 3, and 4, we increase the initial distance between the IVAC and the OH$^-$ group to one, two, and three unit cells, respectively. It can be seen that when the OH$^-$ group is within three unit cells from the vacancy, there is a significant drop in the MSD compared to the IVAC only simulation.  To make sure that we observe a real trend of the MSD slope, and not simply random effects, we repeated several simulations by using different seeds of the random number generator, which correspond to different initial positions and velocities of the iodide ions, and indeed, the trend was confirmed.
This is because the vacancy migration tends to go toward the OH$^-$ group and once it is within one unit cell distance, the migration of the IVAC is blocked.

%\vspace*{-2mm}

\begin{figure}[hbt!]
    \centering
    \includegraphics[width=1.0\linewidth]{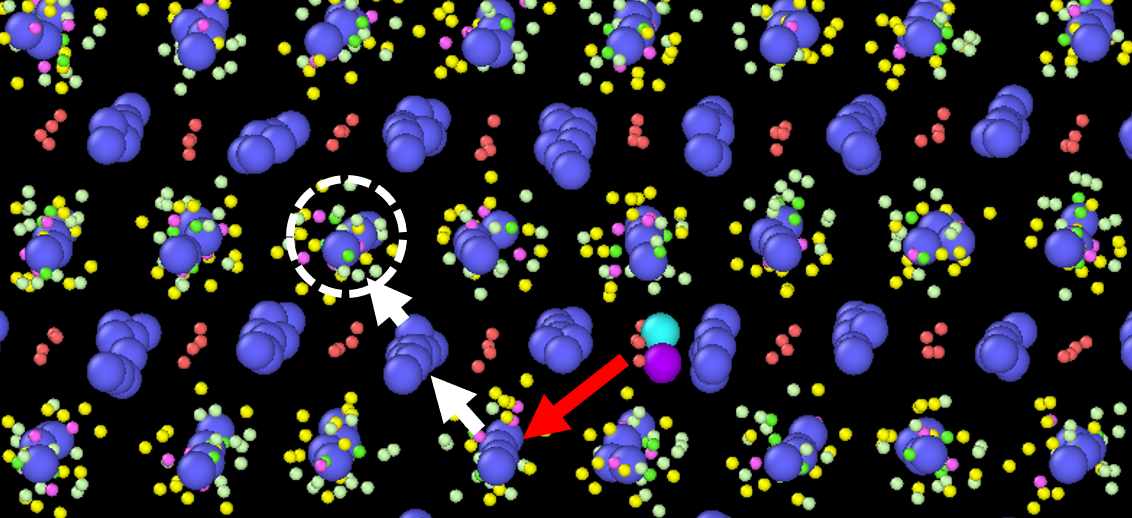}
    \caption{An example of migration path of both the iodides (white arrows) and hydroxyl group (red arrow) in the presence of a vacancy. The initial position of the iodide vacancy is indicated by the dashed circle. The iodides are shown as dark blue spheres, and the OH$^-$ as a light blue-violet group.  }
    \label{fig:migration}
\end{figure}

As we have shown in a previous study \cite{Brophy2023}, when a vacancy defect is placed in the material, another iodide atoms will migrate to fill the vacancy, consequently creating a new vacancy in its place. This process will continue throughout the simulation domain until the vacancy reaches the boundary lines. However, in the presence of the hydroxyl group, the initial migration of iodide atoms will still occur, but, once the newly created vacancy is  within roughly one unit cell from the hydroxyl group, this group will migrate toward the vacancy and will remain very close to it, or possibly fill it itself. Thus, the hydroxyl is capable of stopping the migration of the iodide and allowing the system to return to  stability. Specifically, in Figure \ref{fig:migration}, the white dashed circle indicates the initial placement of the IVAC, and the white arrows show the path of the iodides that have migrated and created new vacancies behind them. The red arrow indicates the migration of the hydroxyl group that has joined with the final vacancy, stopping the migration.

\begin{figure}[hbt!]
    \centering
    \includegraphics[width=1.0\linewidth]{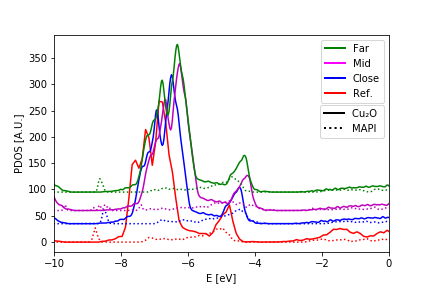}
    \hspace*{-6mm}\includegraphics[width=0.9\linewidth]{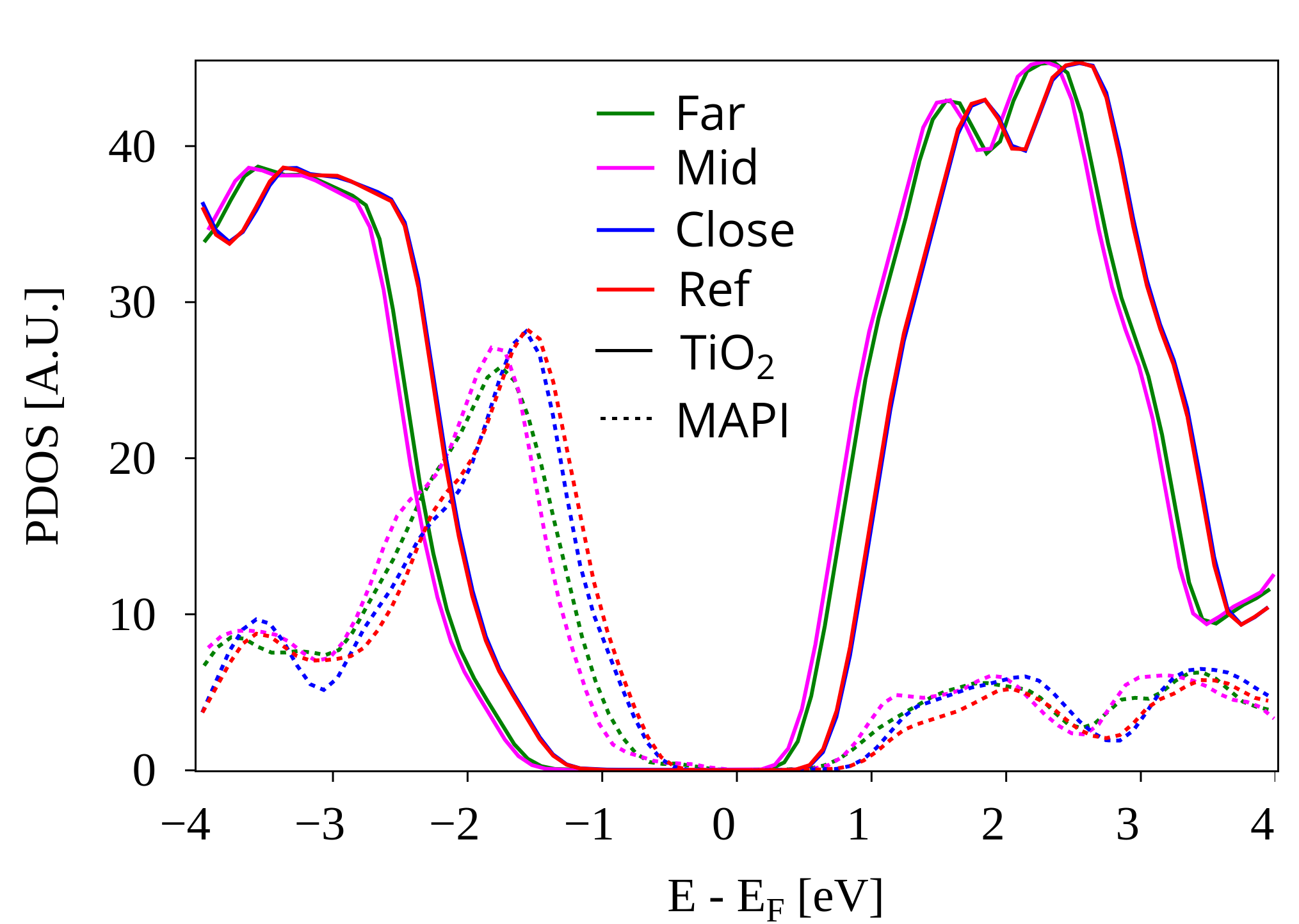}
\vspace*{-2mm}	
 \caption{{\bf Top:} PDOS of MAPI/Cu$_2$O interface: Solid lines represent Cu$_2$O, while dotted lines represent MAPI. The red lines show the ideal MAPI/Cu$_2$O interface, and the blue/green lines have one iodine atom replaced with OH$^{-}$. The blue line represents the interface-close substitution, the green line represents the interface-far substitution, and the magenta line a substitution in the middle. The substitution lines have been shifted upwards for clarity.\\
    {\bf Bottom:} Analog results of the PDOS for MAPI/TiO$_2$ interface. 
    %A qualitatively similar behavior is found for the OH$^-$ substitutions near the MAPI@TiO$_2$ interface, suggesting that their influence on the electronic properties is negligible.
    }
        \label{fig:HO}
        \vspace*{-4mm}
\end{figure}

A natural question is whether the presence of hydroxyl ions in the MAPI does not harm other properties of the solar cell. While only an experimental study can have a definite answer to this question, here we check if the band alignments of MAPI with the electron and hole transporter layers are affected.  

The band alignment at the interface between MAPI and Cu$_2$O, seen as a hole transporter material, was determined using ab initio DFT calculations with SIESTA software. Initially, the band alignment of the ideal interface was computed, and the PDOS for each material was determined for the bulk-like regions, i.e. excluding the layers in the immediate vicinity of the interface and of vacuum, which introduce additional states. Next, a random iodine atom was replaced with OH, and this procedure was repeated 10 times, each time substituting a different iodine. The results for the ideal interface and three cases with OH are presented in the upper part of Figure~\ref{fig:HO}. The calculations showed a slight shift in energy, but the band alignment appeared to be mostly unaffected, and the placement of OH did not have a significant impact on the band alignment. 

We performed similar tests on the MAPI/TiO$_2$, TiO$_2$ being a known electron transporter material for MAPI. Four situations were simulated, with the OH group replacing the iodine atoms, to be compared with the case without iodide vacancies.  A picture qualitatively similar to the case of the hole transporter emerges, as shown in the lower part of Figure~\ref{fig:HO}, again indicating that the band alignment in the presence of (few) OH$^-$ ions remain practically unchanged. Notably, we also found that an uncompensated iodine vacancy shifts the MAPI bands downwards in energy (not shown), meaning that the OH$^-$ can efficiently compensate for this effect. \\

\section{Discussion and conclusion}

% Text from Alex
The binding of OH$^-$ to positively charged iodine vacancy is in principle possible in the context of humid environments. Previously, it was shown that, under ambient conditions, superoxide (O$_2^-$) can be generated at the position of iodine vacancies, which triggers perovskite degradation, leading to the formation of lead iodide, molecular iodine, gaseous methylammonium and water \cite{Aristidou2017}. On the other hand, the formation of OH$^-$ ions can also trigger the deprotonation of the methylammonium, resulting in neutral water and CH$_3$NH$_2$ molecules \cite{Zhang2015}. Therefore, negative ions containing oxygen can passivate iodine vacancies, and likely inhibit iodine migration, but may also prompt the perovskite degradation. However, recent studies have shown that small amounts of coordinating water can have a beneficial effect against degradation, acting as a barrier and preventing the formation of additional hydrates \cite{Contreras2018}. In a similar way, a stability improvement of formamidinium based solar cells was achieved using ambient air additive during fabrication \cite{Salim2021}. Therefore, the potential benefits of the small water amounts should be further investigated experimentally.

The proton H$^+$ resulting from the dissociation of water, being very light,   
should be harmless to MAPI.   Still, it might bind strongly to iodide, neutralizing and blocking the migration of the slow iodide interstitial ions, or eventually can create iodide vacancies which can be immobilized by the OH$^-$. 

In conclusion, we have shown that a negatively charged hydroxyl ion present in MAPI can form a neutral complex with a positively charged iodide vacancy, thereby reducing the migration of the iodide ions via their vacancies. \\

%{\red To be completed ...} {\blue... even if the OH and the vacancy do not meet, the MSD is still decreased slightly when the OH and the vacancy are within roughly 2 unit cells of each other or less, the likelihood of the two meet through migration is increased and this can suppress the negative effects of the vacancy. ..RB these are just thoughts about the conclusion, not in the correct format or worded corrected, just wanted to get my thoughts down}\\

\section*{Acknowledgment}
The research leading to these results has received funding from the EEA Grants 2014–2021, under Project contract no. 36/2021 (project code: EEA-RO-NO-2018-0106).
% and from the Reykjavik University doctoral fund. 
The computational resource was sponsored by EGI and the EGI-ACE H2020 project (GA no. 101017567) with the dedicated support of CLOUDIFIN.\\

\bibliographystyle{ieeetr}
%\bibliography{Bibliography}

%\color{red}
%IEEE conference templates contain guidance text for composing and formatting conference papers. Please ensure that all template text is removed from your conference paper prior to submission to the conference. Failure to remove the template text from your paper may result in your paper not being published.

\end{document}